\documentclass[letter]{aa} 

\usepackage{graphicx,txfonts,natbib,twoopt,breqn}
 \usepackage[breaklinks=true]{hyperref} 
 \bibpunct{(}{)}{;}{a}{}{,}    
 \newcommandtwoopt{\citeads}[3][][]{\href{http://adsabs.harvard.edu/abs/#3}%
                                        {\citealp[#1][#2]{#3}}}
 \newcommandtwoopt{\citepads}[3][][]{\href{http://adsabs.harvard.edu/abs/#3}%
                                        {\citep[#1][#2]{#3}}}
 \newcommandtwoopt{\citetads}[3][][]{\href{http://adsabs.harvard.edu/abs/#3}%
                                        {\citet[#1][#2]{#3}}}
 \newcommandtwoopt{\citeyearads}[3][][]%
   {\href{http://adsabs.harvard.edu/abs/#3}{\citeyear[#1][#2]{#3}}}
\begin{document}

   \title{Detection of photometric variability in the very low-mass binary VHS J1256-1257AB using TESS and Spitzer}
   \titlerunning{Detection of photometric variability in VHS J1256-1257AB using TESS and Spitzer}
   \author{P. A. Miles-P\'aez
          \inst{1}\thanks{ESO Fellow}
          }

   \institute{$^1$European Southern Observatory, Karl-Schwarzschild-Stra{\ss}e 2, 85748 Garching, Germany\\
              \email{pmilespa@eso.org}
             }


 
  \abstract
   {} 
   {We investigate the photometric properties of the M7.5 equal-mass binary VHS J1256-1257AB, which, combined with the late-L dwarf VHS J1256-1257 b, forms one of the few young triple systems of ultra-cool dwarfs currently known.}
   {We analyzed two-minute TESS and two-second {\it Spitzer} archival data with total  durations of about 25 days and 36 hours, respectively. Typical precision in the data is $\pm$1.5\% for TESS and $\pm$0.1\% (in 1 minute) for {\it Spitzer}.  }
   {The optical and infrared light curves periodically exhibit epochs of quasi-sinusoidal modulation followed by epochs of stochastic variability, which resembles the beat pattern created by two waves of similar frequencies that interfere with each other. Our two-wave model for the TESS data shows that the components of VHS J1256-1257AB rotate with periods of $2.0782\pm0.0004$ h and $2.1342\pm0.0003$ h, which is also supported by the {\it Spitzer} observations. As a result, the fluxes of the equally bright VHS J1256-1257A and B alternate between states of phase and anti-phase, explaining the observed photometric variability in their combined light. The projected spectroscopic velocity of VHS J1256-1257AB is remarkably similar to those obtained by combining the measured rotation periods and the expected radii, which indicates that the spin axes of VHS J1256-1257A and B are likely inclined at nearly 90 deg, as previously reported for VHS J1256-1257 b.}
   {}


   \maketitle
%

\section{Introduction}

VHS J1256-1257AB is a 0\farcs1, equal-mass binary with a combined spectral type of M7.5$\,\pm\,$0.5 and an estimated age in the range 150--300 Myr \citep{2015ApJ...804...96G,2016ApJ...818L..12S}. \cite{2020RNAAS...4...54D} reported a trigonometric parallax of 45.0\,$\pm$\,2.4 mas, in agreement with the recent $47.27\pm0.47$ mas from {\it Gaia} Data Release 3 \citep[DR3;][]{2021A&A...649A...1G} that results in a distance of 21.2\,$\pm$\,0.2 pc. The components of the binary have luminosities, $\log{(L_{\rm bol}/L_\odot)}$, of $-2.95\pm0.07$ dex \citep{2020RNAAS...4...54D} that, combined with the age range and the evolutionary models from \cite{2015A&A...577A..42B}, lead to masses and radii in the ranges 0.08--0.11 $M_\odot$ and 0.13--0.15 $R_\odot$, respectively. The binary components are likely to harbor strong magnetic fields in their atmospheres given the known emission in H$\alpha$ \citep{2015ApJ...804...96G} and radio \citep{2018A&A...610A..23G} from the unresolved flux. The system contains a third component at only $\sim$8\arcsec: a very red L dwarf (VHS J1256-1257 b), which has attracted most of the attention in the literature since its  atmosphere resembles those of young giant exoplanets \citep{2020AJ....160...77Z,2020ApJ...893L..30B}. 

We currently know of few triple systems composed of only young dwarfs with spectral types later than M7 \citep[``ultra-cool dwarfs'';][]{2020NatAs...4..650T,2021MNRAS.500.5453S}. Thus, their characterization is important for testing evolutionary models at early ages. In particular, the determination of their individual rotation periods and spectroscopic velocities, combined with their radii, will allow us to learn about the spin-axis relative orientation of the system, which is crucial for putting constraints on their initial conditions of formation and providing insights into angular momentum evolution. In this work we report the detection of photometric variability in both components of VHS J1256-1257AB using archival data of the Transiting Exoplanet Survey Satellite (TESS) and the {\it Spitzer} Space Telescope.
\section{Observations\label{obs}}

\subsection{TESS}
VHS J1256-1257AB ($T$\,=\,$13.346\pm0.021$ mag) was observed with a cadence of 2 minutes by TESS in Sector 10. In this work we use the Pre-search Data Conditioned Simple Aperture Photometry (PDCSAP) light curve provided by the TESS pipeline. We removed any low-quality data point (flagged by the pipeline) as well as any other at more than seven times the standard deviation of the data set. This was the case for less than 1\% of the data. The final light curve consists of 15,743 two-minute exposures, with a median uncertainty of 1.5\%, spanning 24.77 days. The normalized light curve is shown in Fig. \ref{f1} (top), and an image with the TESS field of view and the pipeline aperture is shown in Fig. \ref{a1} of Appendix \ref{A1}. The aperture ($\approx42\arcsec\times42\arcsec$) also covers the positions of VHS J1256-1257 b and another star. However, they are at least 4 mag fainter than VHS J1256-1257AB in the TESS band, and so their flux contribution is negligible.

\begin{figure}
   \centering
   \includegraphics[width=0.4\textwidth]{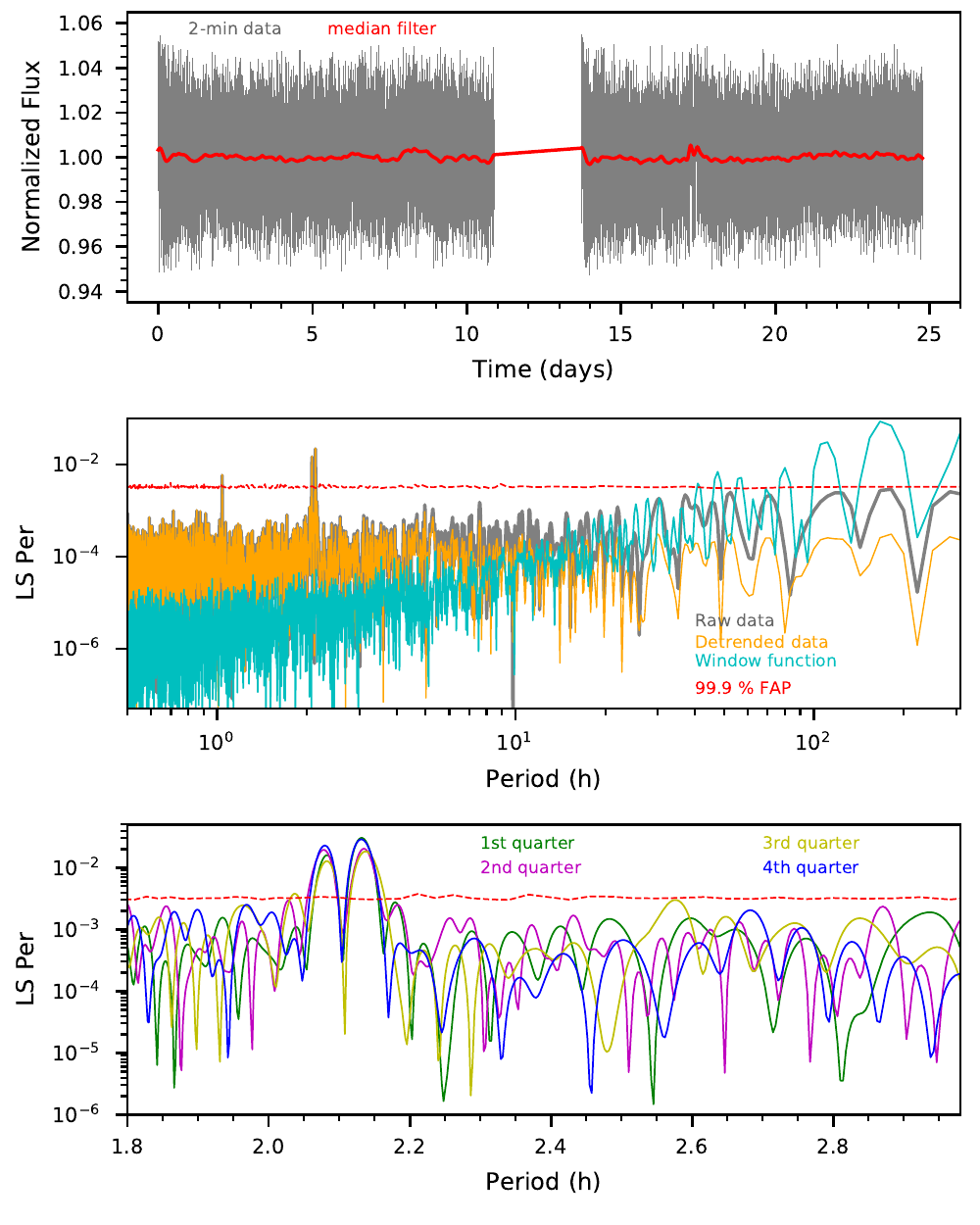}
   \caption{Analysis of the optical data. {\sl Top:} TESS light curve of VHS J1256-1257AB. Vertical lines denote individual uncertainties, and the red line is a 601-point median filter. The zero time corresponds to MJD 58568.93734. {\sl Middle:} LS periodogram for the TESS light curve shown in the top panel  (gray curve) and after detrending with a median filter (orange). The window function is also shown (cyan). We computed the associated 99.9\% FAP (red) from 10,000 light curves generated by data randomization. {\sl Bottom:} LS periodogram for different chunks of the data. A double peak centered at $\sim$2.1 hours is present in all the sections of the data.}
              \label{f1}%
    \end{figure}

\subsection{{\it Spitzer} \label{spt}}
The triple system was also monitored with the {\it Spitzer} Space Telescope using Channel 2 (4.5\,$\mu$m) and the sub-array mode of the IRAC instrument \citep{2004ApJS..154...10F}. Data were collected every 2 seconds over three consecutive 12-hour astronomical observation requests (AORs) and are presented in \citet{2020AJ....160...77Z}. We reanalyzed this data set by obtaining the light curve of VHS J1256-1257AB from the basic calibrated data (BCD) images produced by the {\it Spitzer} Science Center. We extracted the photometry with the {\tt IRAF} task {\tt PHOT} by using a circular aperture with a radius of 2 pixels and subtracted the median value of the sky in a ring with an inner annulus of 5 pixels and a width of 4. We removed all data points lying at more than 7$\sigma$ (typically $<0.3\%$ of the data) and modeled the well-known pixel phase effect in the photometry by fitting a cubic function of the {\sl xy} coordinates \citep{2020AJ....160...38V}:

\begin{dmath}\label{sptcorr}
f(x,y) = A0 + A1\,x + A2\,y + A3\,xy + A4\,x^2 + A5\,y^2 + A6\,x^3 + A7\,y^3 + A8\,x^2y + A9\,xy^2,
\end{dmath}
where $f(x,y)$ stands for the measured flux, $A_i$ are the fitted coefficients, and $x$ and $y$ are the centroid coordinates of VHS J1256-1257AB. The fitting process was done via a Markov chain Monte Carlo (MCMC) using the {\tt emcee} package \citep{2010CAMCS...5...65G,2013PASP..125..306F}. Finally, we divided the measured photometry by the best fit from Eq. \ref{sptcorr}. Our corrected photometry for VHS J1256-1257AB (Fig. \ref{a3}, top) achieved similar results as those presented in \citet[][Fig. 4, right]{2020AJ....160...77Z}: a relatively flat light curve during the first $\sim$24 hours of monitoring and some two-hour periodic variability with amplitudes $<0.1\%$ in the last 12 hours of the data set. Although \citet{2020AJ....160...77Z} could not conclude whether this periodicity had a spurious or astrophysical origin, we investigated whether our correction method could remove some (temporal) variability unrelated to the pixel phase effect. Thus, we modified Eq. \ref{sptcorr} by including a sine function to account for time variability, as done in \citet{2013ApJ...767..173H}:

\begin{dmath}\label{sptcorr2}
F(x,y,t) =  [A0 + A1\,x + A2\,y + A3\,xy + A4\,x^2 + A5\,y^2 + A6\,x^3 + A7\,y^3 + A8\,x^2y + A9\,xy^2]\times [1+B_0\sin{(2\pi\,t/P+\phi)}].
\end{dmath}

\begin{figure}
   \centering
   \includegraphics[width=0.4\textwidth]{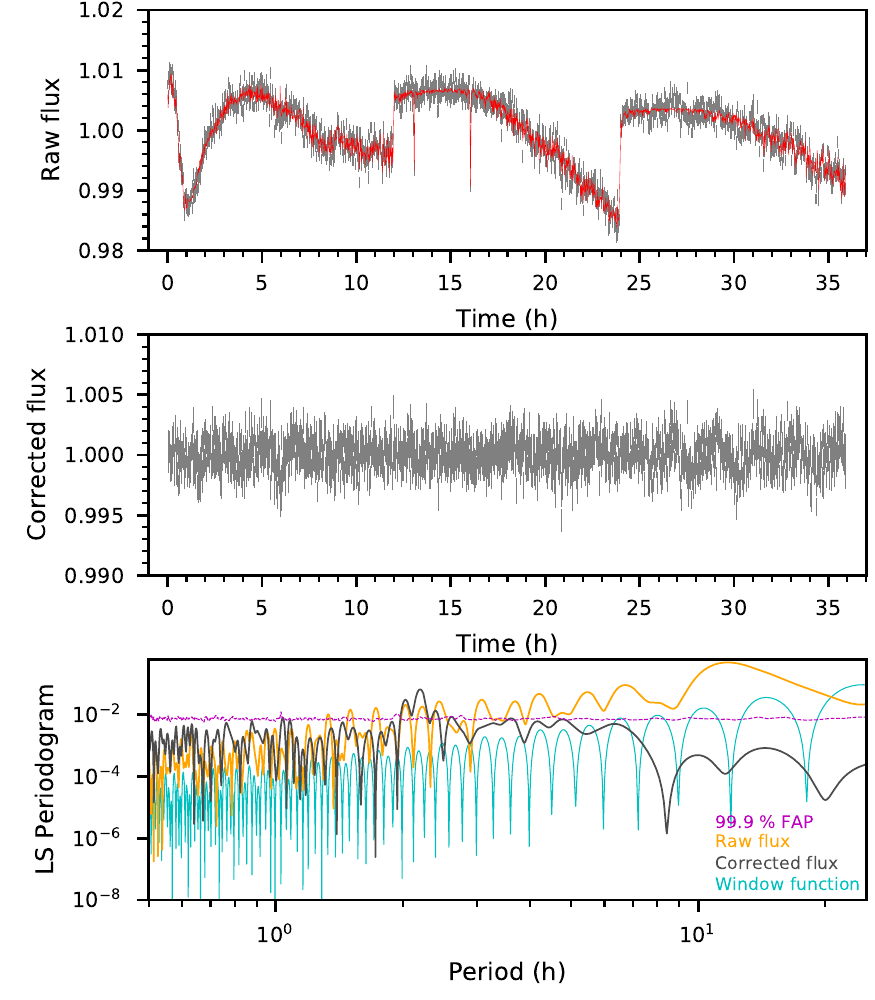}
   \caption{Analysis of the infrared data. {\sl Top:} {\it Spitzer} raw normalized flux of VHS J1256-1257AB as a function of time. The pixel phase model is over-plotted with a red curve. Data are in bins of 32 consecutive two-second images ($\sim$1 minute). Vertical bars indicate the standard deviation of the data in each bin, with a median value of 0.1\,\%. The zero time corresponds to MJD 58801.26092. {\sl Middle:} Pixel-phase-corrected photometry as a function of time. {\sl Bottom:} LS periodogram for the raw and corrected photometry in orange and gray, respectively. The window function (cyan) and the 99.9\% FAP (magenta) are also shown. The pixel-phase-corrected flux also shows a significant peak around $\sim$2.1 h.}
              \label{f2}%
    \end{figure}

We compared the quality of the correction using Eqs. \ref{sptcorr} and  \ref{sptcorr2} by computing the Bayesian information criterion \citep[BIC;][]{1978AnSta...6..461S} for each model and checked their difference ($\Delta{\rm BIC}$). In general, a $\Delta$BIC\,=\,${\rm BIC}_{\rm Model\,1}-{\rm BIC}_{\rm  Model\,2}$$<$2 indicates no significant preference of the data for any model, $\Delta$BIC=2--6 suggests a preference for Model 2, $\Delta$BIC=6--10 is strong evidence against Model 1, and $\Delta{\rm BIC}>10$ indicates that Model 2 is strongly favored by the data. We get a  $\Delta$BIC\,=\,${\rm BIC}_{\rm eq1}-{\rm BIC}_{\rm eq2}\approx3$ for the first two AORs ($\approx$24 h) of data, pointing to no significant difference in correcting our data by Eqs. \ref{sptcorr} or \ref{sptcorr2}. However, we find that the data collected during the last AOR strongly favor Eq. \ref{sptcorr2} as the $\Delta$BIC is $\approx$99. For consistency, we corrected the three AORs with the model from Eq. \ref{sptcorr2}. We used the same $P$ in the three AORs but allowed independent values for $B_0$ and $\phi$, and we find $P=2.15\pm0.03$ h and an amplitude of variability of 0.083$\pm$0.005\% for the last AOR, which is in excellent agreement with those reported by \citet[][0.074$\pm$0.007\%, 2.12$\pm$0.02 h]{2020AJ....160...77Z}. Figure \ref{f2} shows the raw photometry and its best phase-pixel model (top), the corrected photometry (middle), and a Lomb-Scargle (LS) periodogram  \citep[][]{1976Ap&SS..39..447L,1982ApJ...263..835S} of the raw and corrected data (bottom). We also show a comparison of the photometry corrected via Eqs. \ref{sptcorr} and \ref{sptcorr2} in Fig. \ref{a3}, and we show that there is no obvious dependence between the centroid position and the corrected flux in Fig. \ref{a2}. 

\section{Space photometry analysis\label{spa}}

\subsection{TESS shows a two-hour periodicity in VHS J1256-1257AB}

The middle panel of Fig. \ref{f1} shows the LS periodogram of the TESS data of VHS J1256-1257AB (gray) and its associated window function (cyan), which informs on spurious periodicities resulting from the data sampling. We also plot the level for the 99.9\% false alarm probability (FAP; red) that was computed via bootstrapping as in \citet{2017MNRAS.472.2297M} or \citet{2018ApJS..236...16V}. In short, we shuffled the data set, which would destroy any real periodicity, and computed the associated LS periodogram. We repeated this process 10,000 times. Then, for each frequency we investigated the level at which 99.9\% of the simulated LS periodograms showed a smaller value and adopted it as our FAP level. TESS photometry usually exhibits some $>$24-hour trends likely associated with momentum dumps or Moon-Earth scattered light that are not fully removed by the pipeline \citep{2019ESS.....433312V}. We did a first-order correction of these trends by removing from the data a 601-point median filter (red curve in Fig. \ref{f1}, top). The LS periodogram for the detrended data is shown with an orange line in the middle panel of Fig. \ref{f1}. Both data sets show significant peaks at $\sim$2.1 hours (and their likely harmonics at $\sim$1 hour), in agreement with the periodicity seen in the {\it Spitzer} data. We wanted to check whether this 2.1-hour periodicity could arise from only a certain portion of the TESS data set, so we split the data into four parts and computed their LS periodograms (Fig. \ref{f1}, bottom). The significant 2.1-hour periodicity is present in all the TESS data, but it turns out to be double, with centers at $\approx$2.08 h and $\approx$2.14 h. 

\begin{figure}
   \centering
   \includegraphics[width=0.45\textwidth]{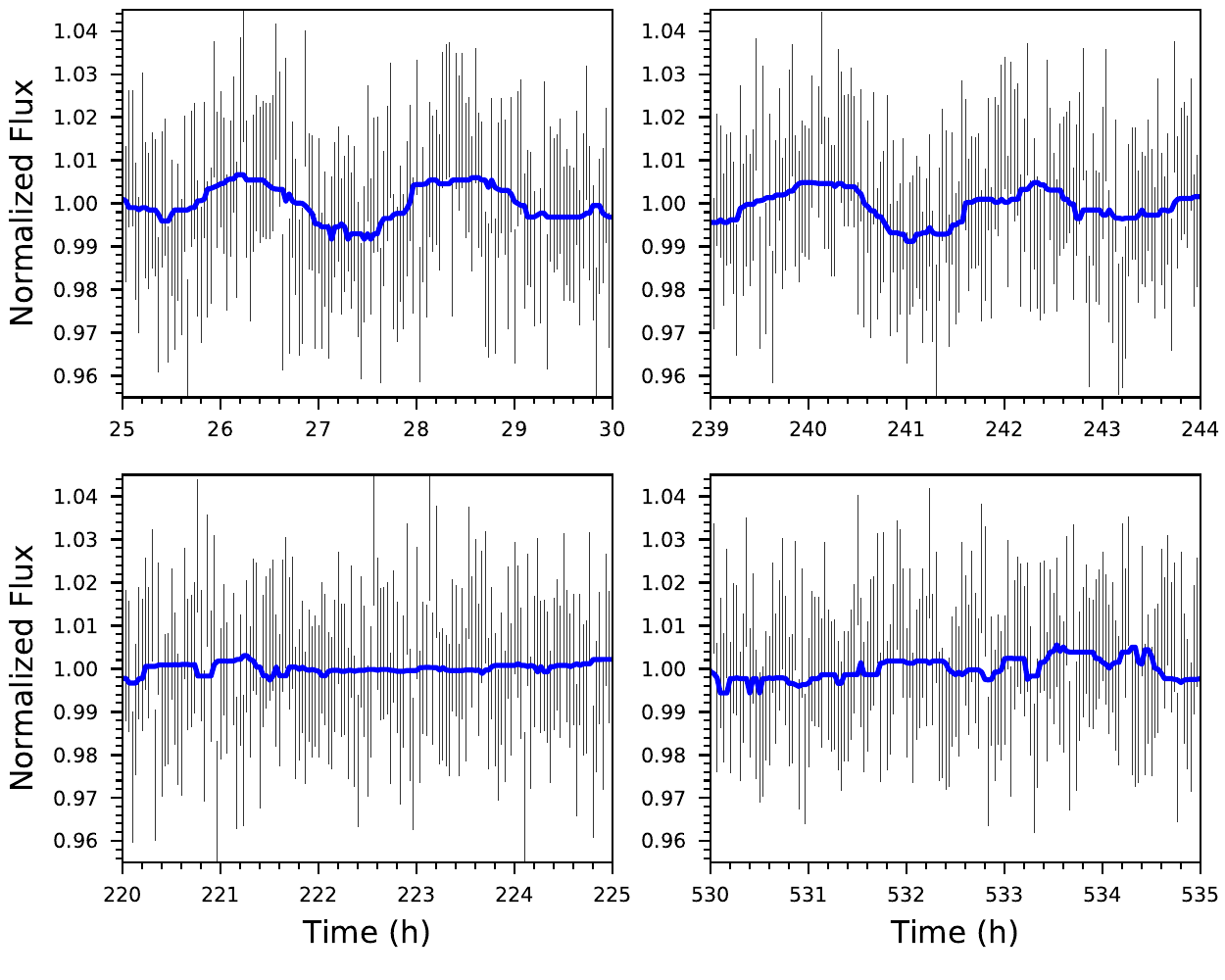}
   \caption{Selected regions of the TESS light curve that show significant photometric modulation with amplitudes of $\sim$0.3\,\% (top) and stochastic variability (bottom). A 31-point median filter is shown in blue.}
              \label{f3}%
    \end{figure}

We took a closer look at the TESS data given the evidence for a 2.1-hour significant periodicity. We find that the light curve exhibits some epochs of quasi-sinusoidal modulation (Fig. \ref{f3}, top panels), followed by epochs of stochastic variability (Fig. \ref{f3}, bottom panels), which resembles the behavior of the {\it Spitzer} data. While the periodicity in the TESS and {\it Spitzer} data is similar, the amplitudes of photometric variability are different (about 0.3\% and 0.1\%  for TESS and {\it Spitzer}, respectively), which is unsurprising as the variability amplitude is generally not expected to be constant across a wide wavelength range.

\subsection{The light from VHS J1256-1257A and B interfere with each other, inducing a beat pattern}

The photometric modulation of VHS J1256-1257AB in two different telescopes and filters strongly supports an astrophysical origin for the observed variability. However, the changing quasi-periodic and stochastic states of  the light curves could also have some contribution from correlated noise. 

\citet{2017MNRAS.466.4250L} presented a Bayesian approach using Gaussian processes (GPs) to compare different models that can explain photometric modulation as either red noise (due to instrumental systematics and/or weather) or as a periodic feature of the star. A GP is defined by two functions. The first one is the "kernel" function that specifies the covariance between the data. The kernel parameters are referred to as hyperparameters and usually lack physical meaning, even though some efforts have been undertaken in this regard \citep[e.g.,][]{2017AJ....154..220F}. The second is the "mean" function, which is determined by the model expected to explain the general behavior of the data. Following \citet{2017MNRAS.466.4250L}, our first GP model ($A$) assumes that the light curve of the target has a constant value $\mu$ (i.e., the mean function) and that the observed modulation can be explained by the covariance of the data using a Mat\'ern-3/2 (M32) kernel \citep{2006gpml.book.....R}, that is parametrized by an amplitude of covariance and a time-scale parameter for the typical duration of the variations. The second model ($B$) also employs a constant value for the mean function of the GP, but uses a ``rotation kernel'' \citep{2017AJ....154..220F} to model the point-to-point covariance. In contrast to the M32 kernel, the rotation kernel incorporates a periodic component with long- and short-term trends that are useful for inferring stellar rotation periods \citep{2020A&A...634A..75B}. We used the {\tt celerite} package  \citep{2017AJ....154..220F} to compute the GP and {\tt emcee} to run the MCMC process that fits models $A$ (non-periodic modulation) and $B$ (quasi-periodic modulation) via 32 walkers with 500 steps of burn-in and 4000 additional steps to sample the space parameter. We used log-uniform priors in the (broad) range (-15,15) for all the hyperparameters of the GP with the exception of the period of the rotation kernel, for which we used a uniform prior with values in the range 1--40 hours, as seen in the field for other ultra-cool dwarfs \citep{2021AJ....161..224T}. We find that both models successfully fit the data. However, the residuals after subtracting the best fit from model $A$ still show a double peak at $\sim$2.1 h in the LS periodogram. On the other hand, the residuals after subtracting model $B$ do not exhibit any periodicity in their associated LS periodogram. We computed the ${\rm BIC}$ of both models and find $\Delta{\rm BIC}=M_{\rm A}-M_{\rm B}\approx182$ and $\Delta{\rm BIC}=108$ for the TESS and {\it Spitzer} data, respectively. Thus, we conclude that the photometric variability of VHS J1256-1257AB (seen in TESS and {\it Spitzer}) is intrinsically periodic rather than the convolution of correlated noise. 

  \begin{figure*}
   \centering
   \includegraphics[width=0.8\textwidth]{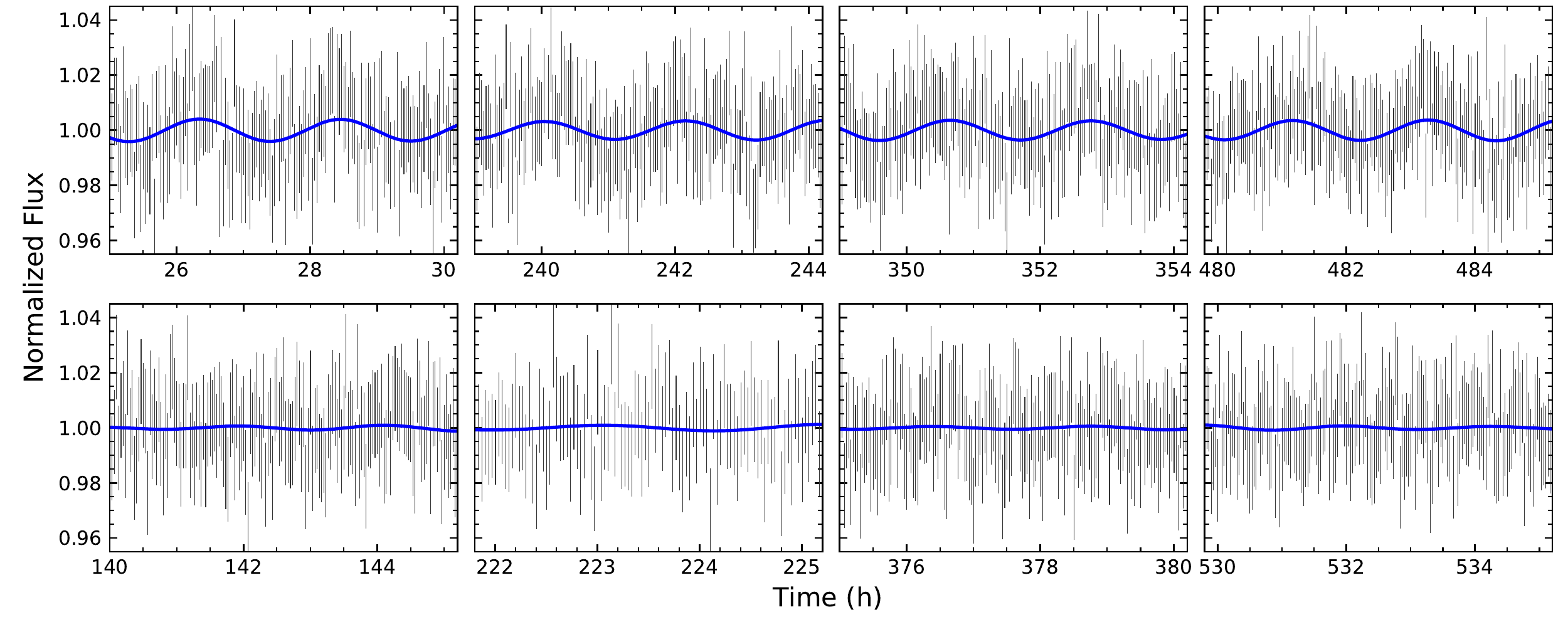}
    \caption{Selected regions of the TESS light curve of VHS J1256-1257AB after removing the best fit for correlated noise from model $E$. The best fit for the interference of  two waves is also plotted in blue. Model $E$ can explain the epochs of modulation (top) and stochastic variability (bottom) in the data. The full data set and the two-wave model are shown in Fig. \ref{a4}.}
              \label{f4}%
    \end{figure*}
    
Model $B$ is an effective model for rotational modulation but cannot provide any other information on the physical parameters of the target apart from the period. Thus, we built a new set of GP models that can explain the systematics in the data but that also have a physical meaning for the periodic variability coming from the light of VHS J1256-1257AB. Table \ref{t1} lists all the models that we explore to explain the modulation seen in the TESS (and {\it Spitzer}) data, as well as the equations for the M32 and the rotation kernels. We now assume that the light curve (modeled by the mean function in Table \ref{t1}) is periodic and has some red noise contamination, which we modeled with an M32 kernel. In model $C$, we investigate whether the light curves can be explained by only a sine function with a red noise component. We find that this model can fit the TESS data reasonably well with a period close to 2.1 h and a $\Delta{\rm BIC}=M_{\rm A}-M_{\rm C}\approx162$. However, the LS periodogram of the residuals still shows the two peaks at $\sim$2.08 and $\sim$2.14 hours. Also, this model can reproduce the modulation seen in Fig. \ref{f3}, but it overestimates the variability in the flat regions. We used a truncated Fourier series of order two (model $D$) to investigate if a better agreement can be achieved, but we find a $\Delta{\rm BIC}=M_{\rm A}-M_{\rm D}\approx135$, which is worse than that of model $C$. Similar results are found for models $C$ and $D$ in the {\it Spitzer} data. Thus, we conclude that a single variable source cannot explain the observed photometric variability.

\begin{figure}
   \centering
   \includegraphics[width=0.45\textwidth]{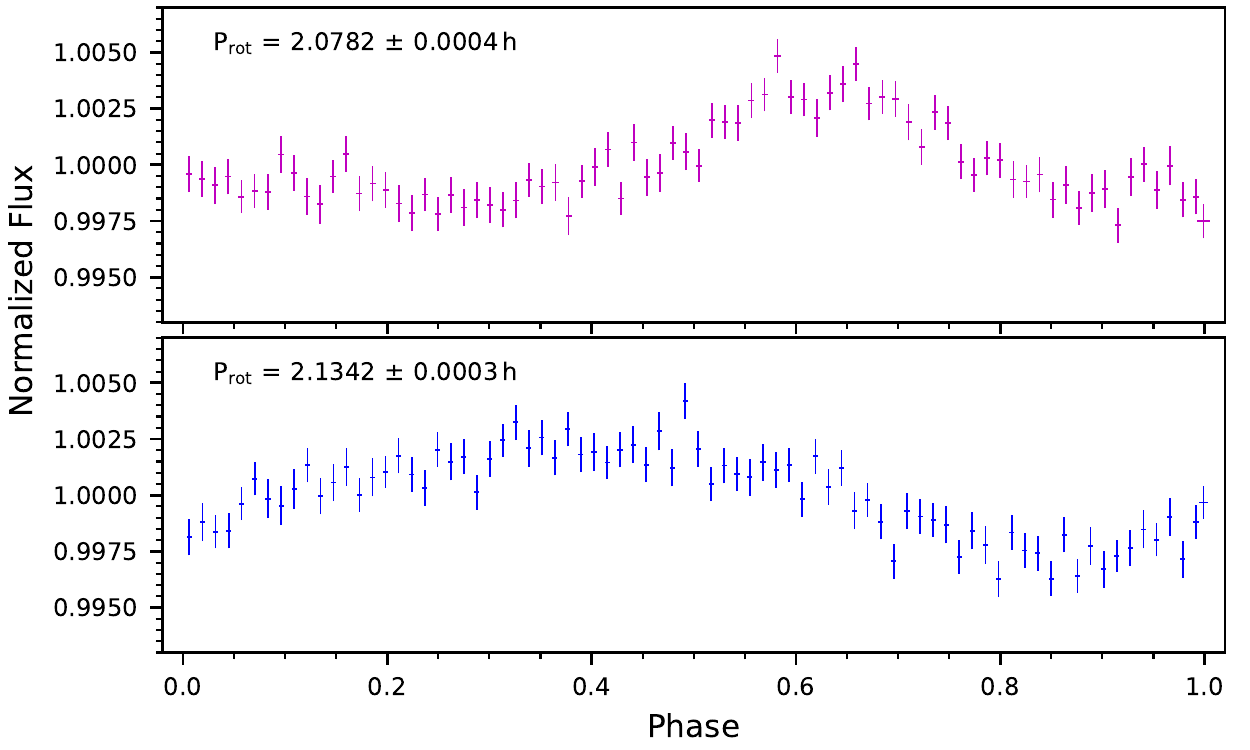}
   \caption{Individual rotation-modulated signals of VHS J1256-1257A and B, shown in  the top and bottom panels, respectively. Each individual signal was computed by removing, from the TESS data, the best fits for the rotation of the other component and the correlated noise. Data for each signal (referred to the same zero time) were phase-folded with their periods and binned every 200 points. Vertical bars indicate the standard deviation of the data in each bin divided by $\sqrt{200}$. Horizontal bars indicate the uncertainty in phase. This was estimated using error propagation and the individual phase uncertainty of the data, computed as $\Delta {\rm Phase}_i\approx\Delta P_{\rm rot}*t_i / P^2_{\rm rot}$.}
              \label{f5}%
    \end{figure}

\begin{figure}
   \centering
   \includegraphics[width=0.45\textwidth]{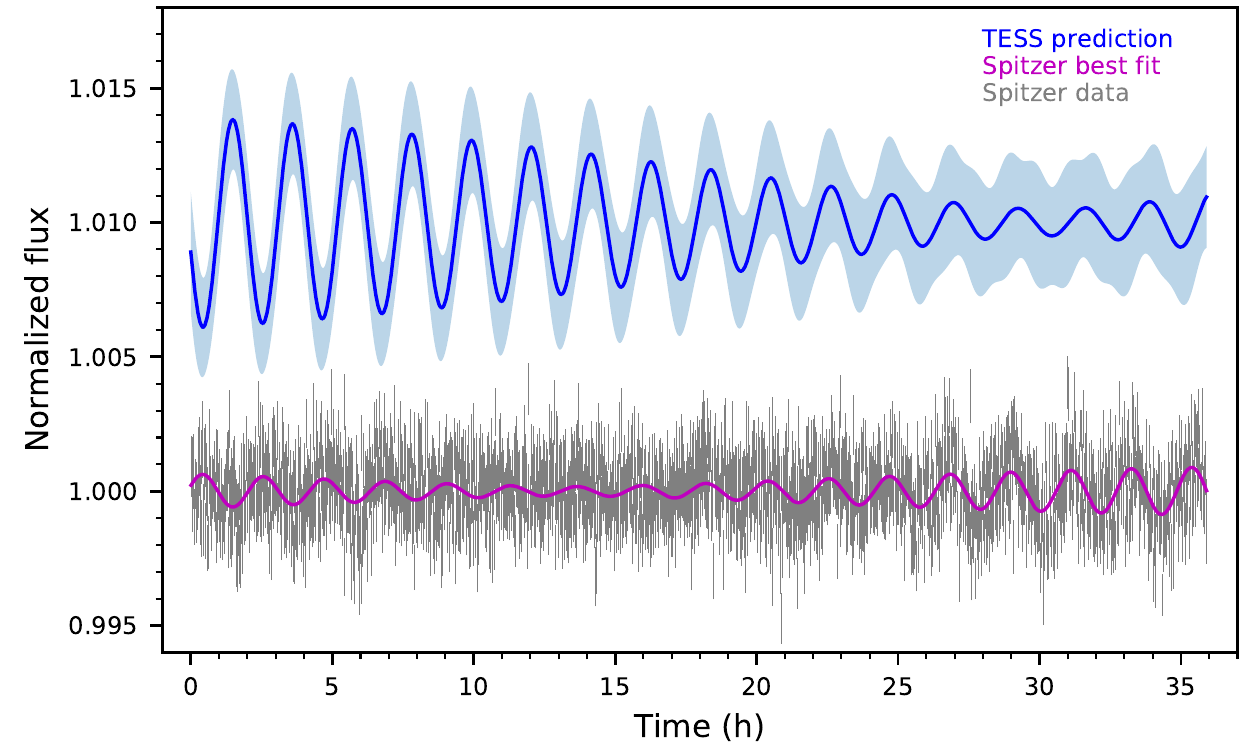}
   \caption{{\it Spitzer} pixel-phase-corrected light curve of VHS J1256-1257AB (gray symbols) and the best fit for model $E$ (magenta). A projection of the TESS variability at the moment of the {\it Spitzer} observations (7.7 months later than the beginning of TESS sector 10) is shown in blue.}
              \label{f6}%
    \end{figure}

The components of VHS J1256-1257AB have the same age and mass, so they should equally imprint their signal in the combined flux measured by TESS and {\it Spitzer}. Thus, in model $E$ we assume that the observed signal is the combination of the light of two periodic sources. We used two sine functions with independent rotation periods and phases, but with the same amplitude for simplicity. In this case we find $\Delta{\rm BIC}=M_{\rm A}-M_{\rm E}\approx687$ and $\Delta{\rm BIC}=M_{\rm B}-M_{\rm E}\approx540$ for the TESS data, showing that the data strongly prefer model $E$ over all the others. The best fit converges to rotation periods of $2.0782\pm0.0004$ h and $2.1342\pm0.0003$ h (very close to the features seen in the periodogram of Fig. 1), an amplitude of photometric variability of $0.21\pm0.02$\,\%, and a difference in phase of $1.6\pm0.2$ rad. In Fig. \ref{f4} we plot the best fit for the two waves and the TESS data corrected from the best fit of the red noise. This model nicely shows how the light of both components of the binary is in phase in some epochs, leading to the quasi-periodic variability seen, and out of phase in other epochs, resulting in stochastic variability. We also generated the individual TESS light curves for each component and show them phase-folded to their periods in Fig. \ref{f5}. Model $E$ represents the classical example of a beat pattern in which two waves with similar frequencies interfere with each other. From the equations for this interference case, the period of the beat pattern (i.e., the modulated variability) is $|1/P_{\rm rot,1}-1/P_{\rm rot,2}|^{-1}=79.2\pm0.7$ h, which explains why only a fraction of the 36-hour {\it Spitzer} data exhibits significant sine-like modulation. The {\it Spitzer} data also show a strong preference for model $E$, with rotation periods of $2.088\pm0.017$ h and $2.163\pm0.010$ h, a phase difference of 4.6$\pm$0.6 rad, and an amplitude of variability of $0.05\pm0.01$\,\%. We note the different uncertainties derived for the rotation periods from TESS and {\it Spitzer}. This is mainly due to the number of consecutive rotation cycles contained in each data set: about 250 and 17 for TESS and {\it Spitzer}, respectively. The longer duration of the TESS data allows us to better constrain the interference pattern and, therefore, the rotation periods. Thus, we adopted the TESS periods for VHS J1256-1257A and B. Figure \ref{f6} shows the best fit for Model $E$ in the {\it Spitzer} data and a projection for the TESS data at the time of the {\it Spitzer} observations, which were collected 232.32 days later. It is not surprising that the TESS and {\it Spitzer} beat patterns are out of phase since the feature responsible for the TESS variability might be different at the moment of the {\it Spitzer} observations. More importantly, the optical and infrared data might prove different physical processes in the atmosphere, as found for late-M dwarfs \citep{2015Natur.523..568H}.

\section{Discussion and conclusions}

Our analysis shows that VHS J1256-1257A and B have almost the same rotation period. This is usually seen in binaries with very short orbital periods that equalize their rotation periods via tidal synchronization \citep{2019ApJ...881...88F}. However, this mechanism seems unlikely in the case of VHS J1256-1257AB as we estimated its orbital period to be $>$4 years by using Kepler's third law with the mass range of the system, the measured distance, and their visual separation. Given that both components have the same mass, age, and (presumably) dynamical history, the similarity in rotation periods seems to be the result of angular momentum evolution. Also, the measured rotation periods and their optical variability are similar to those seen in other late-M dwarfs of similar age and mass \citep{2017MNRAS.472.2297M}.

\cite{2018NatAs...2..138B} determined a $v$\,sin\,$i$ value of 75.2$^{+2.7}_{-2.3}$ km\,s$^{-1}$ for the combined light of VHS J1256-1257AB. When combining the expected radius range of both components and their rotation periods, we find rotational velocities (${\rm v}_{\rm rot}=2\pi\,R/P_{\rm rot}$) in the range 76-87 km\,s$^{-1}$, which suggests that both components likely have spectroscopic velocities close to the reported $v$\,sin\,$i$. \citet{2012ApJ...750...79K} reported $v$\,sin\,$i$ measurements for 11 ultra-cool binaries. They found that these objects are generally fast rotators ($v$\,sin\,$i>$10 km\,s$^{-1}$) with similar $v$\,sin\,$i$ between both components of the binary, and that any difference in $v$\,sin\,$i$ could be explained by differences in mass between the binary components and/or interactions with more massive objects. Thus, we conclude that the individual values of $v$\,sin\,$i$ for VHS J1256-1257A and B should be similar to the one derived for the combined light given the similar rotation periods of both components and the trend delineated by other very low-mass binaries. We derive the spin-axis inclination range of VHS J1256-1257A and B by combining their rotational velocities with the observed $v$\,sin\,$i$ as sin\,$i=v$\,sin\,$i/{\rm v}_{\rm rot}$, and conclude that the spin axis of each component is likely inclined at $>$57 deg, similar to the equator-on view determined for VHS J1256-1257 b by \cite{2020AJ....160...77Z}. \citet{2016ApJ...818L..12S} found that the orbit of VHS J1256-1257A and B cannot be face-on. However, more measurements are needed to determine the orientation of the orbital plane and the likely spin-orbit alignment between VHS J1256-1257A, VHS J1256-1257B, and VHS J1256-1257 b, as seen in more massive systems \citep{1994AJ....107..306H}.

Finally, \citet{2015ApJ...804...96G} reported H$\alpha$ emission in the (unresolved) optical spectrum of the binary, and \citet{2018A&A...610A..23G} found radio emission arising from the same position of VHS J1256-1257AB at X band, which support the presence of strong magnetic fields and magnetic spots as the likely origin of the photometric variability reported here. Silicate clouds are predicted to naturally form at spectral types later than M7 \citep{1997ApJ...480L..39J}, so they could also explain some of the variability. The relative contribution of each mechanism is, however, beyond the scope of this work. We note that the optical light curves of each component (Fig. \ref{f4} and \ref{f5}) seem to be stable over the full duration of the TESS observations (i.e., $\approx$283 rotation cycles). VHS J1256-1257AB will be observed again by TESS in sectors 37 and 47, which will allow us to investigate the stability of the light curves of each component and put them in context with other variable ultra-cool dwarfs.

\begin{table}
\scriptsize
\caption{GP models explored in the TESS and {\it Spitzer} data.}             
\label{t1}      
\centering          
\begin{tabular}{l l c c c}    
\hline\hline       
 \multicolumn{2}{l}{Model} &  \multicolumn{2}{c}{Mean function} & Kernel \\ 
\hline     
  \multicolumn{2}{l}{A: red noise only} &\multicolumn{2}{c}{constant ($\mu$)} &M32 \\        
 \multicolumn{2}{l}{B: quasi-periodic variability} &\multicolumn{2}{c}{constant ($\mu$)} &Rotation \\               
  \multicolumn{2}{l}{C: red noise+single sinusoid} &\multicolumn{2}{c}{$\mu+A_0\,\sin{(2\pi\,t/P_{\rm rot}+\phi)}$} &M32\\      
 \multicolumn{2}{l}{D: red noise + truncated Fourier series} &\multicolumn{2}{c}{$\mu + \sum^{N=2}_{n=1} B_n\,\cos{(2\pi\,n\,t/P_{\rm rot}+\phi_n)}$} &M32\\               
 \multicolumn{2}{l}{E: red noise + double sinusoid} &\multicolumn{2}{c}{$\mu + \sum^{N=2}_{i=1} A_0\,\sin{(2\pi\,t/P_{\rm rot,i}+\phi_i)}$} &M32\\                        
\hline                  
\end{tabular}

\tablefoot{\scriptsize The Mat\'ern-3/2 kernel (M32) is given by the expression $k(r) = a^{2} \left( 1 + \sqrt{ \frac{3r^{2}}{\tau^{2}} } \right) \exp{\left( - \sqrt{\frac{3r^2}{\tau^2}} \right)}$ \citep{2006gpml.book.....R} and the rotation kernel by $k(r) = \dfrac{B}{2+C}~e^{-r/L}\Bigg(\cos\left(\dfrac{2\pi\tau}{P_{rot}}\right)+(1+C)\Bigg)$ \citep{2017AJ....154..220F}, where $r$ is a variable of the time defined as $r=|t_i-t_j|$.}
\end{table}

\begin{acknowledgements}
We are thankful to the anonymous referee for his/her valuable comments. We are also grateful to Dr. Henri M. J. Boffin for useful discussions on rotation in binary stars. This paper includes data collected by the TESS mission. Funding for the TESS mission is provided by the NASA's Science Mission Directorate. This work is also based in part on observations made with the {\it Spitzer} Space Telescope, which was operated by the Jet Propulsion Laboratory, California Institute of Technology under a contract with NASA. This work made use of \texttt{tpfplotter} by J. Lillo-Box (publicly available in www.github.com/jlillo/tpfplotter).
\end{acknowledgements}

%
%

\bibliographystyle{aa} 
\bibliography{biblio.bib} 

\begin{thebibliography}{33}
\expandafter\ifx\csname natexlab\endcsname\relax\def\natexlab#1{#1}\fi

\bibitem[{{Aller} {et~al.}(2020){Aller}, {Lillo-Box}, {Jones}, {Miranda},
  {Barcel{\'o} Forteza}, {Barcel{\'o} Forteza}, \& {Barcel{\'o}
  Forteza}}]{2020A&A...635A.128A}
{Aller}, A., {Lillo-Box}, J., {Jones}, D., {et~al.} 2020, \aap, 635, A128

\bibitem[{{Baraffe} {et~al.}(2015){Baraffe}, {Homeier}, {Allard}, \&
  {Chabrier}}]{2015A&A...577A..42B}
{Baraffe}, I., {Homeier}, D., {Allard}, F., \& {Chabrier}, G. 2015, \aap, 577,
  A42

\bibitem[{{Barros} {et~al.}(2020){Barros}, {Demangeon}, {D{\'\i}az}, {Cabrera},
  {Santos}, {Faria}, \& {Pereira}}]{2020A&A...634A..75B}
{Barros}, S.~C.~C., {Demangeon}, O., {D{\'\i}az}, R.~F., {et~al.} 2020, \aap,
  634, A75

\bibitem[{{Bowler} {et~al.}(2020){Bowler}, {Zhou}, {Morley}, {Kataria},
  {Bryan}, {Benneke}, \& {Batygin}}]{2020ApJ...893L..30B}
{Bowler}, B.~P., {Zhou}, Y., {Morley}, C.~V., {et~al.} 2020, \apjl, 893, L30

\bibitem[{{Bryan} {et~al.}(2018){Bryan}, {Benneke}, {Knutson}, {Batygin},
  {Bowler}, {Bowler}, {Bowler}, \& {Bowler}}]{2018NatAs...2..138B}
{Bryan}, M.~L., {Benneke}, B., {Knutson}, H.~A., {et~al.} 2018, Nature Ast., 2,
  138

\bibitem[{{Dupuy} {et~al.}(2020){Dupuy}, {Liu}, {Magnier}, {Best}, {Baraffe},
  {Chabrier}, {Forveille}, {Metchev}, \& {Tremblin}}]{2020RNAAS...4...54D}
{Dupuy}, T.~J., {Liu}, M.~C., {Magnier}, E.~A., {et~al.} 2020, RNAAS, 4, 54

\bibitem[{{Fazio} {et~al.}(2004){Fazio}, {Hora}, {Allen}, {Ashby}, {Barmby},
  {Deutsch}, {Huang}, {Kleiner}, {Marengo}, {Megeath}, {Melnick}, {Pahre},
  {Patten}, {Polizotti}, {Smith}, {Taylor}, {Wang}, {Willner}, {Hoffmann},
  {Pipher}, {Forrest}, {McMurty}, {McCreight}, {McKelvey}, {McMurray}, {Koch},
  {Moseley}, {Arendt}, {Mentzell}, {Marx}, {Losch}, {Mayman}, {Eichhorn},
  {Krebs}, {Jhabvala}, {Gezari}, {Fixsen}, {Flores}, {Shakoorzadeh}, {Jungo},
  {Hakun}, {Workman}, {Karpati}, {Kichak}, {Whitley}, {Mann}, {Tollestrup},
  {Eisenhardt}, {Stern}, {Gorjian}, {Bhattacharya}, {Carey}, {Nelson},
  {Glaccum}, {Lacy}, {Lowrance}, {Laine}, {Reach}, {Stauffer}, {Surace},
  {Wilson}, {Wright}, {Hoffman}, {Domingo}, \& {Cohen}}]{2004ApJS..154...10F}
{Fazio}, G.~G., {Hora}, J.~L., {Allen}, L.~E., {et~al.} 2004, \apjs, 154, 10

\bibitem[{{Fleming} {et~al.}(2019){Fleming}, {Barnes}, {Davenport}, \&
  {Luger}}]{2019ApJ...881...88F}
{Fleming}, D.~P., {Barnes}, R., {Davenport}, J. R.~A., \& {Luger}, R. 2019,
  \apj, 881, 88

\bibitem[{{Foreman-Mackey} {et~al.}(2017){Foreman-Mackey}, {Agol},
  {Ambikasaran}, {Angus}, {Angus}, {Angus}, \& {Angus}}]{2017AJ....154..220F}
{Foreman-Mackey}, D., {Agol}, E., {Ambikasaran}, S., {et~al.} 2017, \aj, 154,
  220

\bibitem[{{Foreman-Mackey} {et~al.}(2013){Foreman-Mackey}, {Hogg}, {Lang},
  {Goodman}, {Goodman}, {Goodman}, \& {Goodman}}]{2013PASP..125..306F}
{Foreman-Mackey}, D., {Hogg}, D.~W., {Lang}, D., {et~al.} 2013, \pasp, 125, 306

\bibitem[{{Gaia Collaboration} {et~al.}(2021){Gaia Collaboration}, {Brown},
  {Vallenari}, {Prusti}, {de Bruijne}, {Babusiaux}, {Biermann}, {Creevey},
  {Evans}, {Eyer}, {Hutton}, {Jansen}, {Jordi}, {Klioner}, {Lammers},
  {Lindegren}, {Luri}, {Mignard}, {Panem}, {Pourbaix}, {Randich}, {Sartoretti},
  {Soubiran}, {Walton}, {Arenou}, {Bailer-Jones}, {Bastian}, {Cropper},
  {Drimmel}, {Katz}, {Lattanzi}, {van Leeuwen}, {Bakker}, {Cacciari},
  {Casta{\~n}eda}, {De Angeli}, {Ducourant}, {Fabricius}, {Fouesneau},
  {Fr{\'e}mat}, {Guerra}, {Guerrier}, {Guiraud}, {Jean-Antoine Piccolo},
  {Masana}, {Messineo}, {Mowlavi}, {Nicolas}, {Nienartowicz}, {Pailler},
  {Panuzzo}, {Riclet}, {Roux}, {Seabroke}, {Sordo}, {Tanga}, {Th{\'e}venin},
  {Gracia-Abril}, {Portell}, {Teyssier}, {Altmann}, {Andrae}, {Bellas-Velidis},
  {Benson}, {Berthier}, {Blomme}, {Brugaletta}, {Burgess}, {Busso}, {Carry},
  {Cellino}, {Cheek}, {Clementini}, {Damerdji}, {Davidson}, {Delchambre},
  {Dell'Oro}, {Fern{\'a}ndez-Hern{\'a}ndez}, {Galluccio}, {Garc{\'\i}a-Lario},
  {Garcia-Reinaldos}, {Gonz{\'a}lez-N{\'u}{\~n}ez}, {Gosset}, {Haigron},
  {Halbwachs}, {Hambly}, {Harrison}, {Hatzidimitriou}, {Heiter},
  {Hern{\'a}ndez}, {Hestroffer}, {Hodgkin}, {Holl}, {Jan{\ss}en}, {Jevardat de
  Fombelle}, {Jordan}, {Krone-Martins}, {Lanzafame}, {L{\"o}ffler}, {Lorca},
  {Manteiga}, {Marchal}, {Marrese}, {Moitinho}, {Mora}, {Muinonen}, {Osborne},
  {Pancino}, {Pauwels}, {Petit}, {Recio-Blanco}, {Richards}, {Riello},
  {Rimoldini}, {Robin}, {Roegiers}, {Rybizki}, {Sarro}, {Siopis}, {Smith},
  {Sozzetti}, {Ulla}, {Utrilla}, {van Leeuwen}, {van Reeven}, {Abbas}, {Abreu
  Aramburu}, {Accart}, {Aerts}, {Aguado}, {Ajaj}, {Altavilla}, {{\'A}lvarez},
  {{\'A}lvarez Cid-Fuentes}, {Alves}, {Anderson}, {Anglada Varela}, {Antoja},
  {Audard}, {Baines}, {Baker}, {Balaguer-N{\'u}{\~n}ez}, {Balbinot}, {Balog},
  {Barache}, {Barbato}, {Barros}, {Barstow}, {Bartolom{\'e}}, {Bassilana},
  {Bauchet}, {Baudesson-Stella}, {Becciani}, {Bellazzini}, {Bernet}, {Bertone},
  {Bianchi}, {Blanco-Cuaresma}, {Boch}, {Bombrun}, {Bossini}, {Bouquillon},
  {Bragaglia}, {Bramante}, {Breedt}, {Bressan}, {Brouillet}, {Bucciarelli},
  {Burlacu}, {Busonero}, {Butkevich}, {Buzzi}, {Caffau}, {Cancelliere},
  {C{\'a}novas}, {Cantat-Gaudin}, {Carballo}, {Carlucci}, {Carnerero},
  {Carrasco}, {Casamiquela}, {Castellani}, {Castro-Ginard}, {Castro Sampol},
  {Chaoul}, {Charlot}, {Chemin}, {Chiavassa}, {Cioni}, {Comoretto}, {Cooper},
  {Cornez}, {Cowell}, {Crifo}, {Crosta}, {Crowley}, {Dafonte}, {Dapergolas},
  {David}, {David}, {de Laverny}, {De Luise}, {De March}, {De Ridder}, {de
  Souza}, {de Teodoro}, {de Torres}, {del Peloso}, {del Pozo}, {Delbo},
  {Delgado}, {Delgado}, {Delisle}, {Di Matteo}, {Diakite}, {Diener},
  {Distefano}, {Dolding}, {Eappachen}, {Edvardsson}, {Enke}, {Esquej}, {Fabre},
  {Fabrizio}, {Faigler}, {Fedorets}, {Fernique}, {Fienga}, {Figueras},
  {Fouron}, {Fragkoudi}, {Fraile}, {Franke}, {Gai}, {Garabato},
  {Garcia-Gutierrez}, {Garc{\'\i}a-Torres}, {Garofalo}, {Gavras}, {Gerlach},
  {Geyer}, {Giacobbe}, {Gilmore}, {Girona}, {Giuffrida}, {Gomel}, {Gomez},
  {Gonzalez-Santamaria}, {Gonz{\'a}lez-Vidal}, {Granvik},
  {Guti{\'e}rrez-S{\'a}nchez}, {Guy}, {Hauser}, {Haywood}, {Helmi}, {Hidalgo},
  {Hilger}, {H{\l}adczuk}, {Hobbs}, {Holland}, {Huckle}, {Jasniewicz},
  {Jonker}, {Juaristi Campillo}, {Julbe}, {Karbevska}, {Kervella}, {Khanna},
  {Kochoska}, {Kontizas}, {Kordopatis}, {Korn}, {Kostrzewa-Rutkowska},
  {Kruszy{\'n}ska}, {Lambert}, {Lanza}, {Lasne}, {Le Campion}, {Le Fustec},
  {Lebreton}, {Lebzelter}, {Leccia}, {Leclerc}, {Lecoeur-Taibi}, {Liao},
  {Licata}, {Lindstr{\o}m}, {Lister}, {Livanou}, {Lobel}, {Madrero Pardo},
  {Managau}, {Mann}, {Marchant}, {Marconi}, {Marcos Santos}, {Marinoni},
  {Marocco}, {Marshall}, {Martin Polo}, {Mart{\'\i}n-Fleitas}, {Masip},
  {Massari}, {Mastrobuono-Battisti}, {Mazeh}, {McMillan}, {Messina},
  {Michalik}, {Millar}, {Mints}, {Molina}, {Molinaro}, {Moln{\'a}r},
  {Montegriffo}, {Mor}, {Morbidelli}, {Morel}, {Morris}, {Mulone}, {Munoz},
  {Muraveva}, {Murphy}, {Musella}, {Noval}, {Ord{\'e}novic}, {Orr{\`u}},
  {Osinde}, {Pagani}, {Pagano}, {Palaversa}, {Palicio}, {Panahi}, {Pawlak},
  {Pe{\~n}alosa Esteller}, {Penttil{\"a}}, {Piersimoni}, {Pineau}, {Plachy},
  {Plum}, {Poggio}, {Poretti}, {Poujoulet}, {Pr{\v{s}}a}, {Pulone}, {Racero},
  {Ragaini}, {Rainer}, {Raiteri}, {Rambaux}, {Ramos}, {Ramos-Lerate}, {Re
  Fiorentin}, {Regibo}, {Reyl{\'e}}, {Ripepi}, {Riva}, {Rixon}, {Robichon},
  {Robin}, {Roelens}, {Rohrbasser}, {Romero-G{\'o}mez}, {Rowell}, {Royer},
  {Rybicki}, {Sadowski}, {Sagrist{\`a} Sell{\'e}s}, {Sahlmann}, {Salgado},
  {Salguero}, {Samaras}, {Sanchez Gimenez}, {Sanna}, {Santove{\~n}a},
  {Sarasso}, {Schultheis}, {Sciacca}, {Segol}, {Segovia}, {S{\'e}gransan},
  {Semeux}, {Shahaf}, {Siddiqui}, {Siebert}, {Siltala}, {Slezak}, {Smart},
  {Solano}, {Solitro}, {Souami}, {Souchay}, {Spagna}, {Spoto}, {Steele},
  {Steidelm{\"u}ller}, {Stephenson}, {S{\"u}veges}, {Szabados}, {Szegedi-Elek},
  {Taris}, {Tauran}, {Taylor}, {Teixeira}, {Thuillot}, {Tonello}, {Torra},
  {Torra}, {Turon}, {Unger}, {Vaillant}, {van Dillen}, {Vanel}, {Vecchiato},
  {Viala}, {Vicente}, {Voutsinas}, {Weiler}, {Wevers}, {Wyrzykowski}, {Yoldas},
  {Yvard}, {Zhao}, {Zorec}, {Zucker}, {Zurbach}, \&
  {Zwitter}}]{2021A&A...649A...1G}
{Gaia Collaboration}, {Brown}, A.~G.~A., {Vallenari}, A., {et~al.} 2021, \aap,
  649, A1

\bibitem[{{Gauza} {et~al.}(2015){Gauza}, {B{\'e}jar}, {P{\'e}rez-Garrido},
  {Zapatero Osorio}, {Lodieu}, {Rebolo}, {Pall{\'e}}, \&
  {Nowak}}]{2015ApJ...804...96G}
{Gauza}, B., {B{\'e}jar}, V. J.~S., {P{\'e}rez-Garrido}, A., {et~al.} 2015,
  \apj, 804, 96

\bibitem[{{Goodman} \& {Weare}(2010)}]{2010CAMCS...5...65G}
{Goodman}, J. \& {Weare}, J. 2010, Communications in Applied Mathematics and
  Computational Science, 5, 65

\bibitem[{{Guirado} {et~al.}(2018){Guirado}, {Azulay}, {Gauza},
  {P{\'e}rez-Torres}, {Rebolo}, {Climent}, \& {Zapatero
  Osorio}}]{2018A&A...610A..23G}
{Guirado}, J.~C., {Azulay}, R., {Gauza}, B., {et~al.} 2018, \aap, 610, A23

\bibitem[{{Hale}(1994)}]{1994AJ....107..306H}
{Hale}, A. 1994, \aj, 107, 306

\bibitem[{{Hallinan} {et~al.}(2015){Hallinan}, {Littlefair}, {Cotter},
  {Bourke}, {Harding}, {Pineda}, {Butler}, {Golden}, {Basri}, {Doyle}, {Kao},
  {Berdyugina}, {Kuznetsov}, {Rupen}, \& {Antonova}}]{2015Natur.523..568H}
{Hallinan}, G., {Littlefair}, S.~P., {Cotter}, G., {et~al.} 2015, \nat, 523,
  568

\bibitem[{{Heinze} {et~al.}(2013){Heinze}, {Metchev}, {Apai}, {Flateau},
  {Kurtev}, {Marley}, {Radigan}, {Burgasser}, {Artigau}, \&
  {Plavchan}}]{2013ApJ...767..173H}
{Heinze}, A.~N., {Metchev}, S., {Apai}, D., {et~al.} 2013, \apj, 767, 173

\bibitem[{{Jones} \& {Tsuji}(1997)}]{1997ApJ...480L..39J}
{Jones}, H. R.~A. \& {Tsuji}, T. 1997, \apjl, 480, L39

\bibitem[{{Konopacky} {et~al.}(2012){Konopacky}, {Ghez}, {Fabrycky},
  {Macintosh}, {White}, {Barman}, {Rice}, {Hallinan}, \&
  {Duch{\^e}ne}}]{2012ApJ...750...79K}
{Konopacky}, Q.~M., {Ghez}, A.~M., {Fabrycky}, D.~C., {et~al.} 2012, \apj, 750,
  79

\bibitem[{{Littlefair} {et~al.}(2017){Littlefair}, {Burningham}, \&
  {Helling}}]{2017MNRAS.466.4250L}
{Littlefair}, S.~P., {Burningham}, B., \& {Helling}, C. 2017, \mnras, 466, 4250

\bibitem[{{Lomb}(1976)}]{1976Ap&SS..39..447L}
{Lomb}, N.~R. 1976, \apss, 39, 447

\bibitem[{{Miles-P{\'a}ez} {et~al.}(2017){Miles-P{\'a}ez}, {Pall{\'e}}, \&
  {Zapatero Osorio}}]{2017MNRAS.472.2297M}
{Miles-P{\'a}ez}, P.~A., {Pall{\'e}}, E., \& {Zapatero Osorio}, M.~R. 2017,
  \mnras, 472, 2297

\bibitem[{{Rasmussen} \& {Williams}(2006)}]{2006gpml.book.....R}
{Rasmussen}, C.~E. \& {Williams}, C. K.~I. 2006, {Gaussian Processes for
  Machine Learning}

\bibitem[{{Sahlmann} {et~al.}(2021){Sahlmann}, {Dupuy}, {Burgasser},
  {Filippazzo}, {Mart{\'\i}n}, {Bardalez Gagliuffi}, {Hsu}, {Lazorenko}, \&
  {Liu}}]{2021MNRAS.500.5453S}
{Sahlmann}, J., {Dupuy}, T.~J., {Burgasser}, A.~J., {et~al.} 2021, \mnras, 500,
  5453

\bibitem[{{Scargle}(1982)}]{1982ApJ...263..835S}
{Scargle}, J.~D. 1982, \apj, 263, 835

\bibitem[{{Schwarz}(1978)}]{1978AnSta...6..461S}
{Schwarz}, G. 1978, Annals of Statistics, 6, 461

\bibitem[{{Stone} {et~al.}(2016){Stone}, {Skemer}, {Kratter}, {Dupuy}, {Close},
  {Eisner}, {Fortney}, {Hinz}, {Males}, {Morley}, {Morzinski}, \&
  {Ward-Duong}}]{2016ApJ...818L..12S}
{Stone}, J.~M., {Skemer}, A.~J., {Kratter}, K.~M., {et~al.} 2016, \apjl, 818,
  L12

\bibitem[{{Tannock} {et~al.}(2021){Tannock}, {Metchev}, {Heinze},
  {Miles-P{\'a}ez}, {Gagn{\'e}}, {Burgasser}, {Marley}, {Apai}, {Su{\'a}rez},
  \& {Plavchan}}]{2021AJ....161..224T}
{Tannock}, M.~E., {Metchev}, S., {Heinze}, A., {et~al.} 2021, \aj, 161, 224

\bibitem[{{Triaud} {et~al.}(2020){Triaud}, {Burgasser}, {Burdanov}, {Kunovac
  Hod{\v{z}}i{\'c}}, {Alonso}, {Bardalez Gagliuffi}, {Delrez}, {Demory}, {de
  Wit}, {Ducrot}, {Hessman}, {Husser}, {Jehin}, {Pedersen}, {Queloz},
  {McCormac}, {Murray}, {Sebastian}, {Thompson}, {Van Grootel}, \&
  {Gillon}}]{2020NatAs...4..650T}
{Triaud}, A. H.~M.~J., {Burgasser}, A.~J., {Burdanov}, A., {et~al.} 2020,
  NatAst., 4, 650

\bibitem[{{VanderPlas}(2018)}]{2018ApJS..236...16V}
{VanderPlas}, J.~T. 2018, \apjs, 236, 16

\bibitem[{{Vanderspek}(2019)}]{2019ESS.....433312V}
{Vanderspek}, R. 2019, in AAS/Division for Extreme Solar Systems Abstracts,
  Vol.~51, AAS/Division for Extreme Solar Systems Abstracts, 333.12

\bibitem[{{Vos} {et~al.}(2020){Vos}, {Biller}, {Allers}, {Faherty}, {Liu},
  {Metchev}, {Eriksson}, {Manjavacas}, {Dupuy}, {Janson}, {Radigan-Hoffman},
  {Crossfield}, {Bonnefoy}, {Best}, {Homeier}, {Schlieder}, {Brandner},
  {Henning}, {Bonavita}, \& {Buenzli}}]{2020AJ....160...38V}
{Vos}, J.~M., {Biller}, B.~A., {Allers}, K.~N., {et~al.} 2020, \aj, 160, 38

\bibitem[{{Zhou} {et~al.}(2020){Zhou}, {Bowler}, {Morley}, {Apai}, {Kataria},
  {Bryan}, {Benneke}, {Benneke}, {Benneke}, \& {Benneke}}]{2020AJ....160...77Z}
{Zhou}, Y., {Bowler}, B.~P., {Morley}, C.~V., {et~al.} 2020, \aj, 160, 77

\end{thebibliography}

\begin{appendix}
\section{Additional figures \label{A1}}

Here we provide supporting material for the TESS and {\it Spitzer} light curves presented in Sect. \ref{obs} and \ref{spa}.

 \begin{figure}
   \centering
   \includegraphics[width=0.48\textwidth]{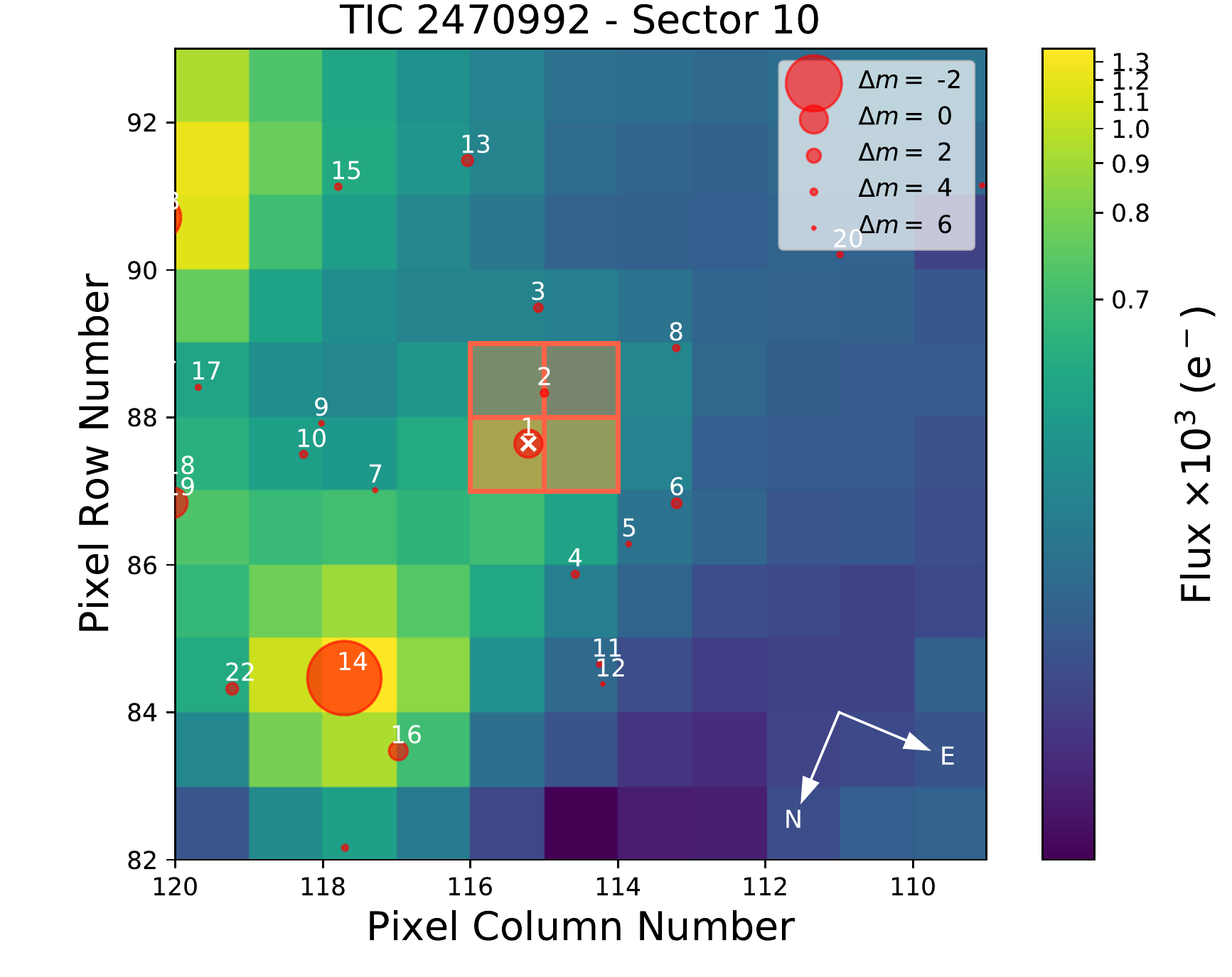}
   \caption{TESS image of the field of VHS J1256-1257AB in Sector 10 \citep[created with {\tt tpfplotter}][]{2020A&A...635A.128A}. The pixels outlined in red indicate the aperture used to extract the light curve shown in the top panel of Fig. \ref{f1}. The size of the red circles indicates the TESS magnitudes of all nearby stars. Each TESS pixel covers an area of $\approx21\arcsec\times21\arcsec$. The aperture used includes the signal of VHS J1256-1257AB as well as those of VHS J1256-1257 b and a star with {\it Gaia} ID 3526198180427873664. VHS J1256-1257AB is at least 4 mag brighter than any other source in the aperture.}
              \label{a1}%
    \end{figure}

 \begin{figure}
   \centering
   \includegraphics[width=0.48\textwidth]{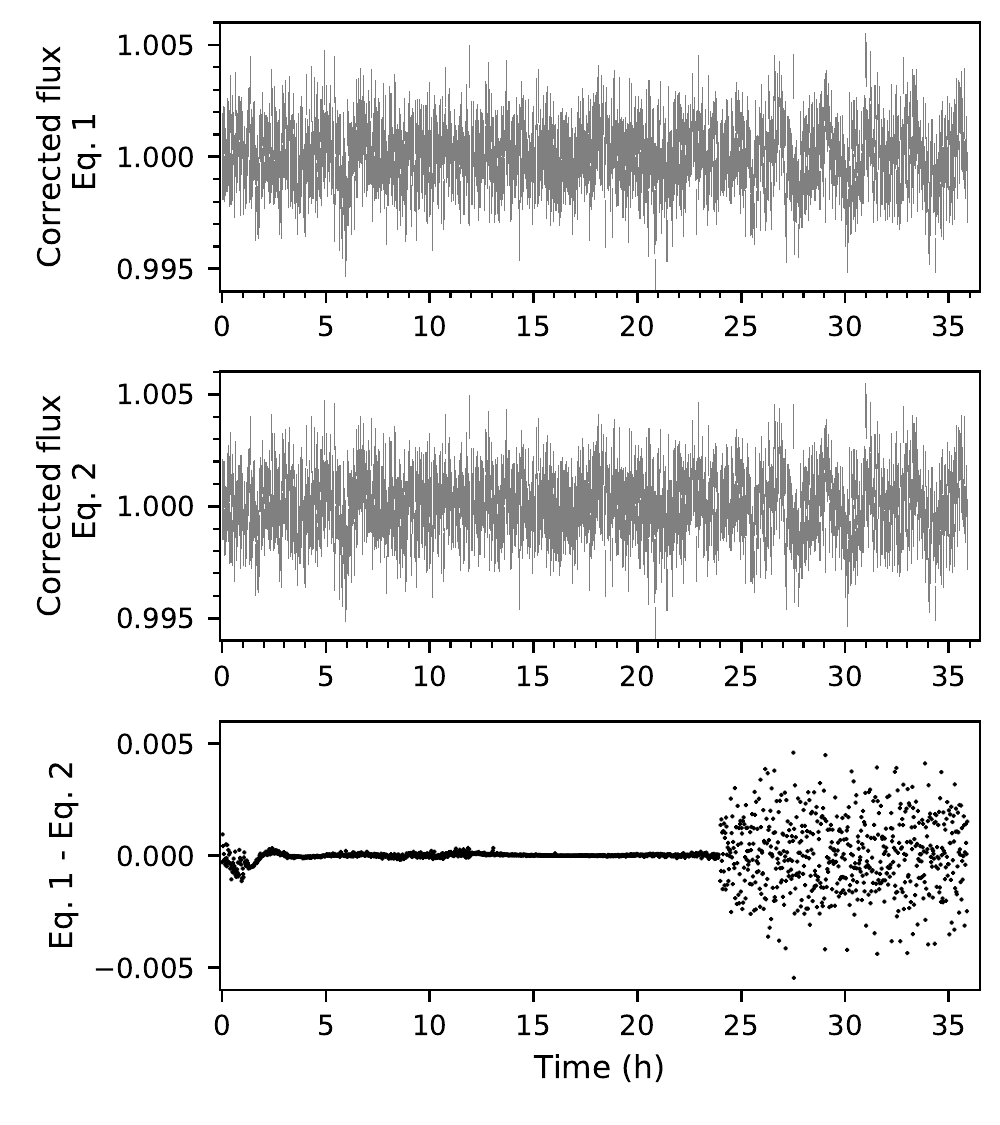}
   \caption{Comparison of the {\it Spitzer} pixel-phase-corrected flux using Eq. \ref{sptcorr} (top) and Eq. \ref{sptcorr2} (middle). Data have been averaged every 32 images ($\approx$1 min) and normalized. Vertical bars stand for the standard deviation of each one-minute bin. Both solutions seem rather alike, even though the BIC comparison favors Eq. \ref{sptcorr2}. The bottom panel shows the difference of the corrected data shown in the top and middle panels. The first $\approx$24 h of data do not show a significant modulation, so the amplitude of the periodic signal in Eq. \ref{sptcorr2} tends to zero, and so Eq. \ref{sptcorr2}\,$\approx$\,Eq.\ref{sptcorr}.
}
              \label{a3}%
    \end{figure}

 \begin{figure}
   \centering
   \includegraphics[width=0.48\textwidth]{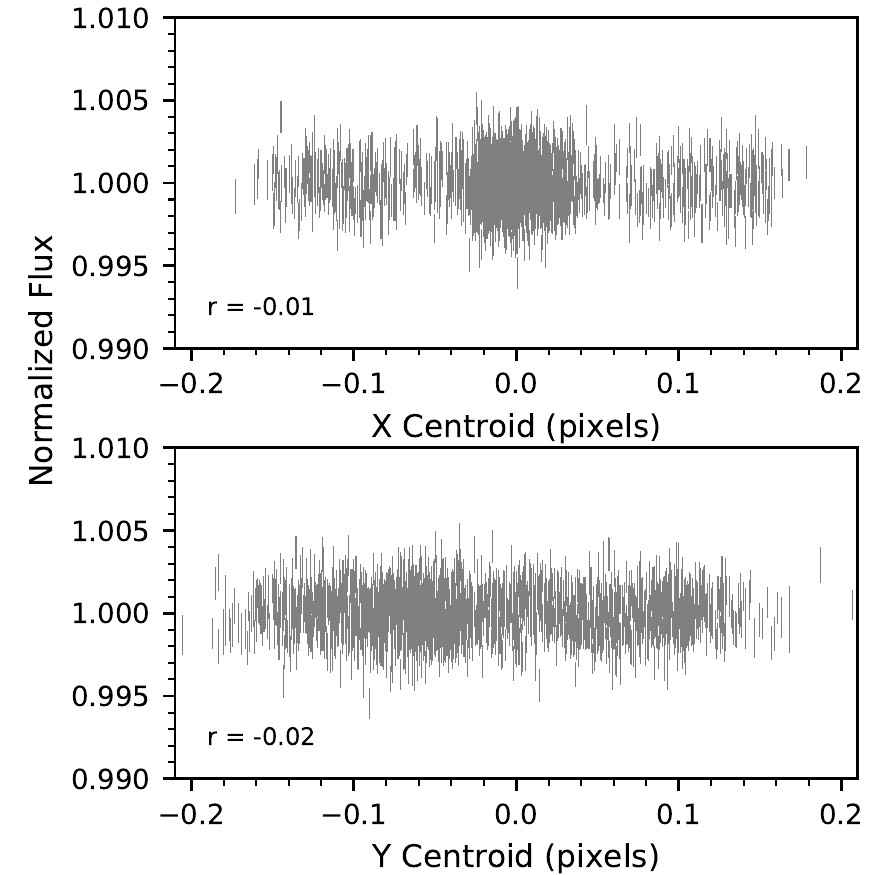}
   \caption{{\it Spitzer} pixel-phase-corrected flux as a function of centroid position in both the x and y directions. The centroids are measured relative to the average centroids across all exposures. We also compute the Pearson correlation coefficient (r) in each panel and find that there is no correlation between the corrected flux and the target's $xy$ position.
}
              \label{a2}%
    \end{figure}

 \begin{figure*}
   \centering
   \includegraphics[angle=90,origin=c,width=0.85\textwidth]{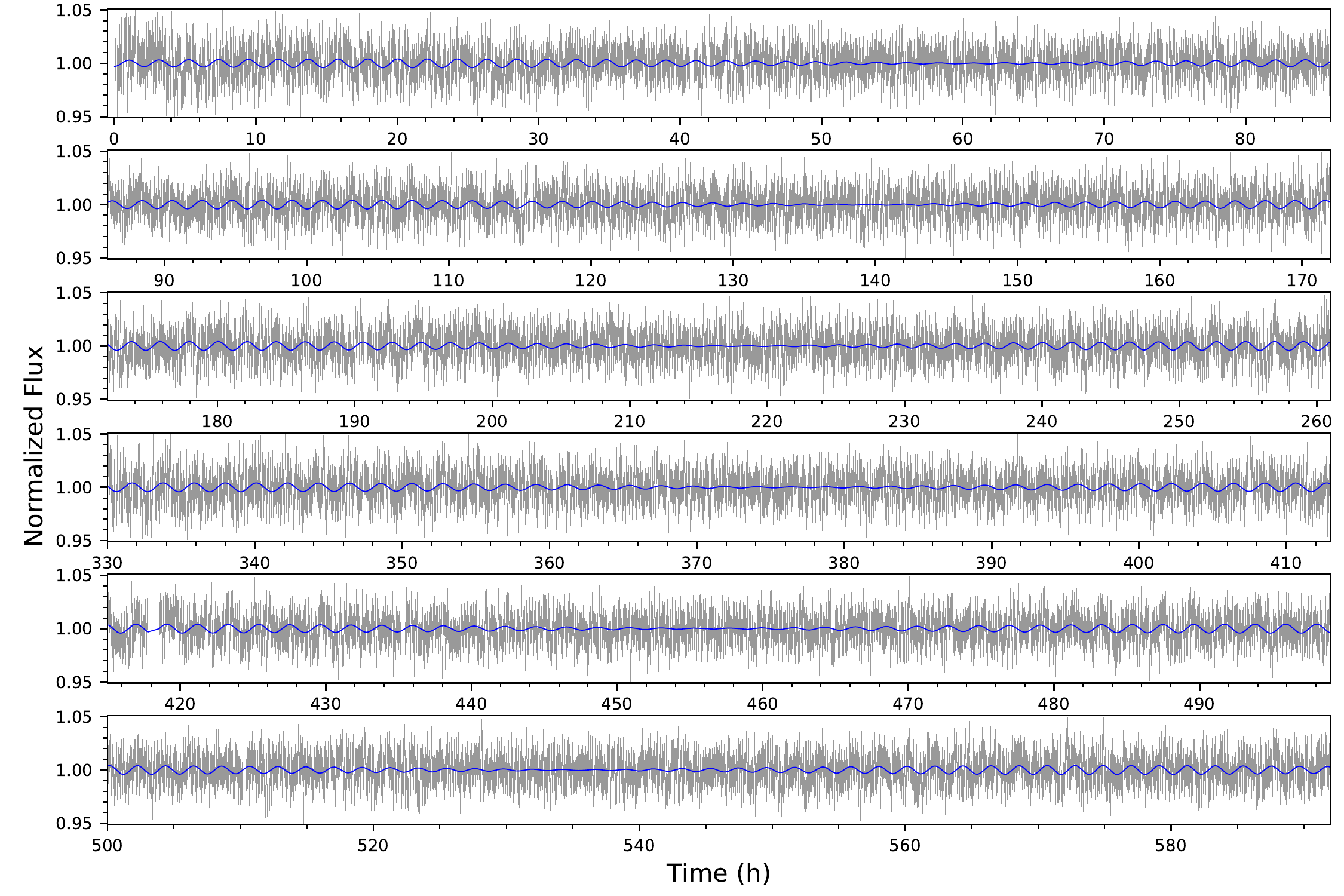}
    \caption{TESS light curve of VHS J1256-1257AB after removing the best fit for correlated noise from model $E$. The best fit for the interference of  two waves is plotted in blue. Model $E$ can explain the epochs of modulation and stochastic variability in the data.}
              \label{a4}%
    \end{figure*}

\end{appendix}

\end{document}